\documentstyle[multicol,aps,psfig]{revtex}
\begin{document}

\title{Non-Abelian Energy Loss at Finite Opacity}

\author{M. Gyulassy$^{1}$, P. Levai$^{1,2}$, and I. Vitev$^{1}$ }

\address{$^1$  Dept. Physics, Columbia University, 
       538 W 120-th Street, New York, NY 10027, USA\\
$^2$ KFKI Research Institute for Particle and Nuclear Physics, PO Box 49,
      Budapest, 1525, Hungary}

\maketitle

\begin{abstract}
A systematic expansion in opacity, $L/\lambda$,
is used to clarify the non-linear behavior of induced gluon
radiation in quark-gluon  plasmas.  The inclusive
differential gluon distribution is calculated up to
second order in opacity and compared to the zeroth order (factorization)
limit.  The opacity expansion makes it possible
to take finite kinematic constraints into account that suppress
jet quenching in nuclear collisions
below RHIC ($\sqrt{s}=200$~AGeV) energies.
\vspace{.2cm}

\end{abstract}

\begin{multicols}{2} 

{\em Introduction.} The production
of high transverse momentum jets in QCD
is always accompanied by 
gluon showers.
For jets  produced inside  nuclei via $e+A$ 
or quark-gluon plasmas via $A+A$,
final state interactions of jet and radiated gluons induce
further radiation and also broaden the gluon shower.
The non-abelian radiative energy loss in a medium is expected to be observable
as ``jet quenching''.  The observable consequences
in nuclear collisions should be seen as a suppression of the 
high $p_\perp$ tails 
of single hadron distributions~\cite{gptw,mgxw92}
and a broadening of the jet cone~\cite{bdms_cone}.

Non-abelian energy loss in pQCD can be calculated analytically in
two limits. In both cases, the energy of the leading parton 
is assumed to be large enough that its  angular deflection can be neglected.
One analytic limit applies to thin plasmas where the mean number, 
$\bar{n}=L/\lambda$, of jet scatterings is small~\cite{mgxw,glv1b}.
 The other limit is the thick plasma one~\cite{bdms,zah,urs},
 where $\bar{n}\gg 1$.
This mean number is a measure of the opacity or
geometrical thickness of the medium:
\begin{equation}
\bar{n}= \frac{N\sigma_{el}}{A_\perp}
=\int dz \, (\sigma_{el}\rho) \approx 
 \, \frac{dN}{dy} \, \frac{\sigma_{el}}{2 \pi R_G^2}
 \, \log \frac{R_G}{\tau_0}  \;.
\label{opac}
\end{equation}
For a constant (box) density $\rho=N/(L A_\perp)$, $L$ is
the target thickness and $\lambda=1/(\rho \sigma_{el})$ is the average
mean free path. For a sharp cylinder geometry we can interpret $L=1.2\, A^{1/3}
\equiv R_s$ as the nuclear radius. 
For a more realistic  3+1D expanding Gaussian cylinder
with $\rho({\bf x}_\perp,\tau)=(\tau_0/\tau)\,\rho_0 \exp \left(
-({\bf x}_\perp^2+\Delta \tau^2)/R_G^2\right) 
I_0 (2|{\bf x}_\perp| \Delta \tau/R_G^2) $,
 $L$ is replaced by the equivalent rms Gaussian transverse radius
$R_G=0.75\, A^{1/3}$~fm.  The rightmost form in (\ref{opac})
is obtained by averaging
over the expanding Gaussian cylinder. Here 
$\tau_0$  is the formation time of the plasma. 

At RHIC energies ($\sqrt{s}\sim200$~AGeV) the  expected rapidity
 density of the gluons
is $dN/dy \sim 1000$ for  $A=200$. With an elastic cross section
$\sigma_{el}\sim 2$~mb and a plasma formation time $\sim 0.5$~fm/c,
we obtain ${\bar{n}} \sim 4$. This suggests
 that neither analytic approaches
may be strictly applicable in practice.
However, because of the non-abelian analog~\cite{mgxw,bdms} 
of the Landau-Pomeranchuk-Migdal (LPM) effect, 
the radiation intensity angular distribution and total energy loss are 
controlled by the combined effect of the number of scatterings
$\bar{n}=L/\lambda$ and  a  formation probability
$p_f= L/l_f = L \mu^2/(2xE)$ of the gluon in the medium. 
In~\cite{bdms,zah} the
induced energy loss was shown to be nonlinear
in the nuclear thickness, $\Delta E\propto \mu^2 L^2/\lambda$
assuming the $\bar{n}\gg 1$ limit.  In~\cite{glv1b}, we showed
that the
angular pattern was even more strongly nonlinear
in the exclusive  (tagged target) case.
We show below that in the inclusive case,
important features of the radiation 
pattern are in fact  governed by the product 
$\bar{n} p_f= L^2(\mu^2/k^+ \lambda)$.
Because of this,  even the first order in opacity
reproduces the $L^2$ dependence of the energy loss
 and the range of applicability of the  finite
opacity expansion derived below may extend to realistic 
targets. 

Another important motivation for our approach is the apparent absence
of jet quenching observed at SPS energies~\cite{XNWA98,MGPL}.
As we find below, the opacity expansion shows that finite kinematic
constraints suppress greatly the non-abelian energy loss at SPS energies.
The work reported in this Letter  extends Ref.~\cite{glv1b} 
by including virtual corrections~\cite{bdms}
necessary to compute inclusive (vs. tagged) jet quenching.
Detailed derivation of these results is given in Ref.~\cite{glv2}.

{\em Hard Gluon Distribution.} At zeroth order in opacity,
the gluon emission from  the  hard production vertex  
is approximately  given by  
\begin{equation}  
x\frac{dN^{(0)}}{dx\, d {\bf k}^2_\perp}= 
 \frac{C_R \alpha_s}{\pi}
\left( 1-x+\frac{x^2}{2}\right)
\frac{1}{{\bf k}^2_\perp} \;\;, 
\label{hdist}
\end{equation}
where 
$x=k^+/E^+ \approx \omega/E$, and $C_R$ is the Casimir 
of the (spin $1/2$) jet in the $d_R$ dimensional color representation.
For a spin 1 jet the gluon splitting function must be substituted above.
The differential energy distribution outside a cone defined by
${\bf k}^2_{\perp}>\mu^2$ is given by
\begin{equation} 
\frac{dI^{(0)}}{dx} = \frac{2 C_R \alpha_s}{\pi} 
\left( 1-x+\frac{x^2}{2}\right)
\, E \, \log \frac{|{\bf k}_\perp|_{\rm max}}{\mu} \;,    
\label{di0}
\end{equation} 
where the upper kinematic limit is
\begin{equation}
\quad {\bf k}^2_{\perp \, \max}=\min\, [4E^2x^2,4E^2x(1-x)]\;.
\label{klimits}
\end{equation}
The energy loss outside the cone is then given by
\begin{equation}
\Delta E^{(0)}=\frac{4C_R\alpha_s}{3\pi}\, E \,
\log \frac{E}{\mu}
\label{de0}
\end{equation}
in the leading $\log(E/\mu)$ approximation (LLA).
While this overestimates the 
radiative  energy loss in the vacuum (self-quenching), it is important to note
that $\Delta E^{(0)}/E \sim 50\%$ is typically much larger
than the medium induced energy loss.

{\em Opacity Expansion.}
To compute the induced radiation,
 we assume as in \cite{mgxw,glv1b} that  the quark gluon plasma 
can be modeled 
by $N$ well-separated, i.e. $\lambda\gg 1/\mu$,  color-screened 
Yukawa potentials. In contrast to~\cite{glv1b}, 
we consider here the {\em inclusive}
gluon distribution induced by a finite medium 
which remains {\em unobserved}.
Without target tagging, we must add
double Born (Virtual) amplitudes $V_i$ to the real (Direct) amplitudes
$D_i$ derived in~\cite{glv1b}.
We denote by $G_0$ the basic eikonal hard emission amplitude 
\begin{equation}
G_0 =  - 2 i g_s \, {\bf \epsilon}_\perp \cdot {\bf H}\;
e^{i  \omega_0 z_0}  c \;, 
\label{g0}
\end{equation}
where we use the shorthand notation of~\cite{glv1b}, 
${\bf H} \equiv {\bf k}_\perp /{\bf k}^2_\perp,\; 
\omega_0 \equiv  {\bf k}^2_\perp/2xE$. The hard parton originates 
at $z_0$, and $c$ denotes the color matrix $T_c$ in its
$d_R$ dimensional representation. 
In Ref.~\cite{glv2} we  developed an iterative procedure 
to generate from this amplitude 
the sum of all amplitudes to {\em any} order in opacity
including both  direct and virtual terms.

To first order in opacity it is 
sufficient to consider the classes of amplitudes denoted $G_0D_1$ 
(sum of three graphs) and 
$G_0V_1$ (sum of four graphs) resulting from  direct and virtual
terms due to a scattering  at position~$z_1$.
The cross section for the induced  radiation  consists of $3^2$ 
real and $2\times 4$ double Born contributions  
that sum to a simple result 
\begin{eqnarray}
&& \, {\rm Tr} \left( G_0D_1D_1^\dagger G_0^\dagger + 
( G_0 V_1 G_0^\dagger + {\rm h.c.}) \right)   \nonumber \\[.5ex]
&&\quad = 4 g_s^2 \, C_R C_A d_R \, \left[ \,
( - 2\,{\bf C}_1 \cdot {\bf B}_1 ) \,
\,\left( 1- \cos ( \omega_1 \Delta z_1 )\right) \, \right] \;,
\label{dn1amps}
\end{eqnarray}
with ${\bf C}_m$ and $\omega_m$ obtained from    ${\bf H}$ and $\omega_0$
through the substitution ${\bf k}_\perp \Rightarrow 
{\bf k}_\perp - {\bf q}_{\perp m}$ and ${\bf B}_m \equiv 
{\bf H}-{\bf C}_m$. Note that the interaction of the radiated
gluon in the medium brought in
a factor $C_A=N_c$.
 Unlike the tagged case~\cite{glv1b}, the above first
order correction to the hard (factorization) distribution in Eq.(\ref{hdist})
has no simple classical cascade limit. In the tagged case,
the result contained terms with  color factors
$C_R^2 d_R$ as well as $C_R C_A  d_R$ that were easily interpretable
in terms of jet rescattering and induced radiation.
In the inclusive case, the virtual corrections lead to long range non-local
interference effects that have no classical analog.
On the other hand, a remarkable ``color triviality" \cite{ursnew}
is found at the inclusive level where all jet rescattering effects 
cancel~\cite{bdms}. As show in detail in Ref.~\cite{glv2},
 the color factor at order $n=2$ is simply $C_R C_A^2  d_R$,
in contrast to the vastly more complicated color structure 
of the tagged case~\cite{glv1b}.

The non-abelian LPM effect is seen in Eq.(\ref{dn1amps})
as arising from the gluon formation factor
\begin{equation}
\Phi(\Delta z_1)= 1- \cos ( \omega_1 \Delta z_1 )\;\;,
\label{phase}
\end{equation}
where $\Delta z_1 =z_1-z_0$. This must also be 
averaged over the longitudinal
target profile, $n(z)$. For a box density~\cite{bdms} of thickness 
$L= 1.2\, A^{1/3}$, $n=\theta(L-z)/L$. 
For analytic simplicity we take instead an exponential
form $n_e(z)=e^{-z/L_e}/L_e$. In order to compare results for the two 
cases, we must require identical mean target depths, i.e. $L_e=L/2$.
With $n_e(z)$ the ensemble averaged formation factor is
\begin{eqnarray}
&&\int_0^\infty \frac{2\, d \Delta z_1}{L} 
e^{-\frac{2\Delta z_1}{L}} \Phi(\Delta z_1)
= \frac{({\bf k} - {\bf q}_1)_\perp^4  L^2}
{16x^2E^2 +({\bf k} - {\bf q}_1)_\perp^4 L^2 } \;. 
\nonumber \\[.5ex]          
\label{lorentz}
\end{eqnarray}
The formation probability in this case is controlled by 
simple Lorentzian factors.         

Averaging over the momentum transfer ${\bf q_{1\perp}}$ via the color
Yukawa potential leads finally
to the gluon double differential distribution
\begin{eqnarray}
x\frac{dN^{(1)}}{dx\, d {\bf k}^2_\perp}&=& 
x\frac{dN^{(0)}}{dx\, d {\bf k}^2_\perp} 
\, \frac{L}{\lambda_g}  \int_0^{q_{\max}^2} d^2{\bf q}_{1\perp} \, 
\frac{ \mu_{eff}^2 }{\pi ({\bf q}_{1\perp}^2 + \mu^2)^2 }
\nonumber \\[.5ex]
&& \qquad \qquad \qquad \frac{ 2\,{\bf k}_\perp \cdot {\bf q}_{1\perp}
  ({\bf k} - {\bf q}_1)_\perp^2  L^2}
{16x^2E^2 +({\bf k} - {\bf q}_1)_\perp^4  L^2 } \;\;.          
\label{dnx1}
\end{eqnarray}
where the opacity factor $L/\lambda_g=N\sigma_{el}^{(g)}/A_\perp$ 
arises from the sum over the $N$ distinct
targets. Note that the radiated gluon mean free path
$\lambda_g=(C_A/C_R)\lambda$ appears rather than the jet mean free path. 
It is the color triviality of Eq.~(\ref{dn1amps})
that  allows  us to absorb $C_R$ factor into
the hard distribution, Eq.~(\ref{hdist}), and $C_A$ factor into $\lambda_g$.

The upper kinematic bound on the momentum transfer
 $q^2_{\rm max}= s/4 \simeq 3 E \mu$, ($1/\mu_{eff}^2=1/\mu^2-
1/(\mu^2+q_{\max}^2)$). For SPS and RHIC energies,
this finite limit cannot be ignored as we show below.

The  second order contribution in opacity
requires a more complex
calculation involving the sum of  $7^2$ direct and $2\times 96$ virtual 
terms as discussed in \cite{glv2}. The result is
\begin{eqnarray}
&& {\rm Tr} 
\left( \, G_0D_1D_2D_2^\dagger D_1^\dagger G_0^\dagger +  
( G_0 V_1D_2D_2^\dagger G_0^\dagger + {\rm h.c.})  \right. 
\nonumber \\[.5ex]
&& \left. \quad 
+\,( G_0D_1V_2 D_1^\dagger G_0^\dagger + {\rm h.c.})  + 
( G_0 V_1 V_2 G_0^\dagger + {\rm h.c.} ) \,  \right) 
\nonumber \\[.5ex]
&& =\, 4 g_s^2 \, C_R C_A^2 d_R \, \left[ \, 
2\,{\bf C}_1 \cdot {\bf B}_1 \,
\left( 1- \cos ( \omega_1 \Delta z_1 )\right) \right.  
\nonumber \\[.8ex]
&&\quad +\, 2\, {\bf C}_2 \cdot {\bf B}_2 \,
\left( \cos ( \omega_2 \Delta z_2 ) - 
\cos ( \omega_2 (\Delta z_1 + \Delta z_2 )\right) 
\nonumber \\[.8ex]
&&\quad- \,2\, {\bf C}_{(12)} \cdot {\bf B}_2 \,
\left( \cos ( \omega_2 \Delta z_2 )
 - \cos ( \omega_{(12)} \Delta z_1 + \omega_2 \Delta z_2 )\right) 
\nonumber \\[.8ex]
&&\left. \quad -   \,2\, {\bf C}_{(12)} \cdot {\bf B}_{2(12)} \,
\left( 1- \cos ( \omega_{(12)} \Delta z_1 )\right) \, \right]\;, 
\label{secord}
\end{eqnarray} 
where with ${\bf C}_{(mn)}$ and $\omega_{(mn)}$ obtained from    
${\bf H}$ and $\omega_0$
through the substitution ${\bf k}_\perp \Rightarrow 
{\bf k}_\perp - {\bf q}_{\perp m}-{\bf q}_{\perp n} $ and 
${\bf B}_{m(nl)} \equiv  {\bf C}_m - {\bf C}_{(nl)}$.
Note the color triviality of this inclusive
result\cite{bdms,ursnew}
 in  contrast to the tagged target case in~\cite{glv1b}.
Both Eqs.~(\ref{dn1amps},\ref{secord}) actually hold for either 
quark or gluon jets in the small $x$ approximation.

We note further that in the two important limits 
$|{\bf k_\perp}| \rightarrow 0$ 
and $|{\bf k_\perp}| \rightarrow \infty$ the distributions vanish for
any fixed $x$ due to azimuthal angular averaging
\begin{equation}
x\frac{dN^{(n)}}{dx\, d {\bf k}^2_\perp} \propto
\int_0^{2\pi} \frac{d\phi}{2\pi}\,
 {\bf k}_\perp \cdot {\bf q}_{m\perp}\, = \, 0 \;\;.
\label{phiav}
\end{equation} 
This is clearly seen in Fig.~1.

The average over the scattering points, $z_1,z_2$ in the second order case
is performed with the exponential density profile
as follows:
\begin{equation}
\langle \, \cdots \, \rangle = \int_0^\infty\frac{3\, d\Delta z_1}{L}
\int_0^\infty\frac{3\,d\Delta z_2}{L} 
e^{-\frac{3 (\Delta z_1+\Delta z_2)}{L}} \, \cdots \;,
\end{equation}
where $\Delta z_i=z_i-z_{i-1}$.
We must use $L_e=L/(n+1)$ for n$^{\rm th}$ order in opacity
in order to insure that 
the first moment of the m$^{\rm th}$ scattering center is identical
to that in a box distribution ($\langle z_m\rangle_n= m L/(n+1)$)
as discussed in~\cite{glv1b}.

Numerical results comparing the first and second order in opacity corrections
to the hard distribution Eq.~(\ref{hdist}) are illustrated in Figs.~1 and 2. 
We consider a 50~GeV quark jet in a medium with $\lambda_g=1$~fm. A screening
scale $\mu=0.5$~GeV and $\alpha_s=0.3$ were assumed. The ``angular"
 distribution in Fig.~1 shows that the first order
angular distribution is wider than the medium independent
hard ($1/{\bf k}_\perp^2$) distribution. The second order corrections 
redistributes the gluons further.
In Fig.~2 the ${\bf k_\perp}$ integrated contributions
to the gluon intensity,
$dI/dx$, are shown for the same conditions as in Fig.~1.
The induced intensity is concentrated at  small $x$ in contrast
 to the relatively constant intensity originating from the hard 
``self-quenching" term~(\ref{di0}). The long $1/x$ tail contributes
however a logarithmic factor $\log(E/\mu)$ as we discuss further below.
The second order correction suppresses the intensity at small $x$ and 
enhances it somewhat at higher
$x \geq 0.2\,$. We note that
 these two effects tend to cancel
in the integrated energy loss as seen in Fig.~3.

{\em Radiation Intensity and Energy loss.}
To gain analytic insight into the
above numerical results, we consider 
the first order induced radiation intensity 
$dI^{(1)}/dx$ in the approximation that ${\bf k}^2_{\perp \, \max}=\infty$.
This allows us to change variables  
${\bf q}_\perp^{\;\prime}\equiv {\bf k}_\perp-{\bf q}_{1\perp}$ 
in Eq.~(\ref{dnx1}) and express the integrand in the azimuthal 
$\phi$ integral as a partial derivative with respect to 
${\bf k}^2_\perp$.  The remaining  
${\bf q}_\perp^{\;\prime}$
integral can be performed then analytically, resulting in
\begin{equation}
\frac{dI^{(1)}}{dx} =\frac{C_R\alpha_s}{\pi} \,
\left( 1 - x + \frac{x^2}{2} \right) \,E \, 
\frac{L}{\lambda_g}\, f(\gamma,\delta) \;,
\label{didx1}
\end{equation} 
where $\gamma={L\mu^2}/({4xE})$ is a formation
probability and $\delta=\mu/E$ is a measure of the importance
of finite kinematics.
The formation function is given by
\begin{equation}
f\left(\gamma,\delta\right) = 
\frac{\gamma \,\left(2\tan^{-1}\frac{\gamma}{\delta}
+\gamma\, \log\frac{\gamma^2 +\delta^2}{1+\delta^2} \right)}
{(1+\gamma^2)} \;,
\label{fapprox}
\end{equation}
which in the $\delta\rightarrow 0$ and $\gamma\ll 1$ limit
reduces to a simple form 
$f(\gamma,0) \; \approx \; \pi \gamma\propto L$.
It is this latter limit that leads to the characteristic quadratic
dependence on $L$:
\begin{center}
\vspace*{8.5cm}
\includegraphics{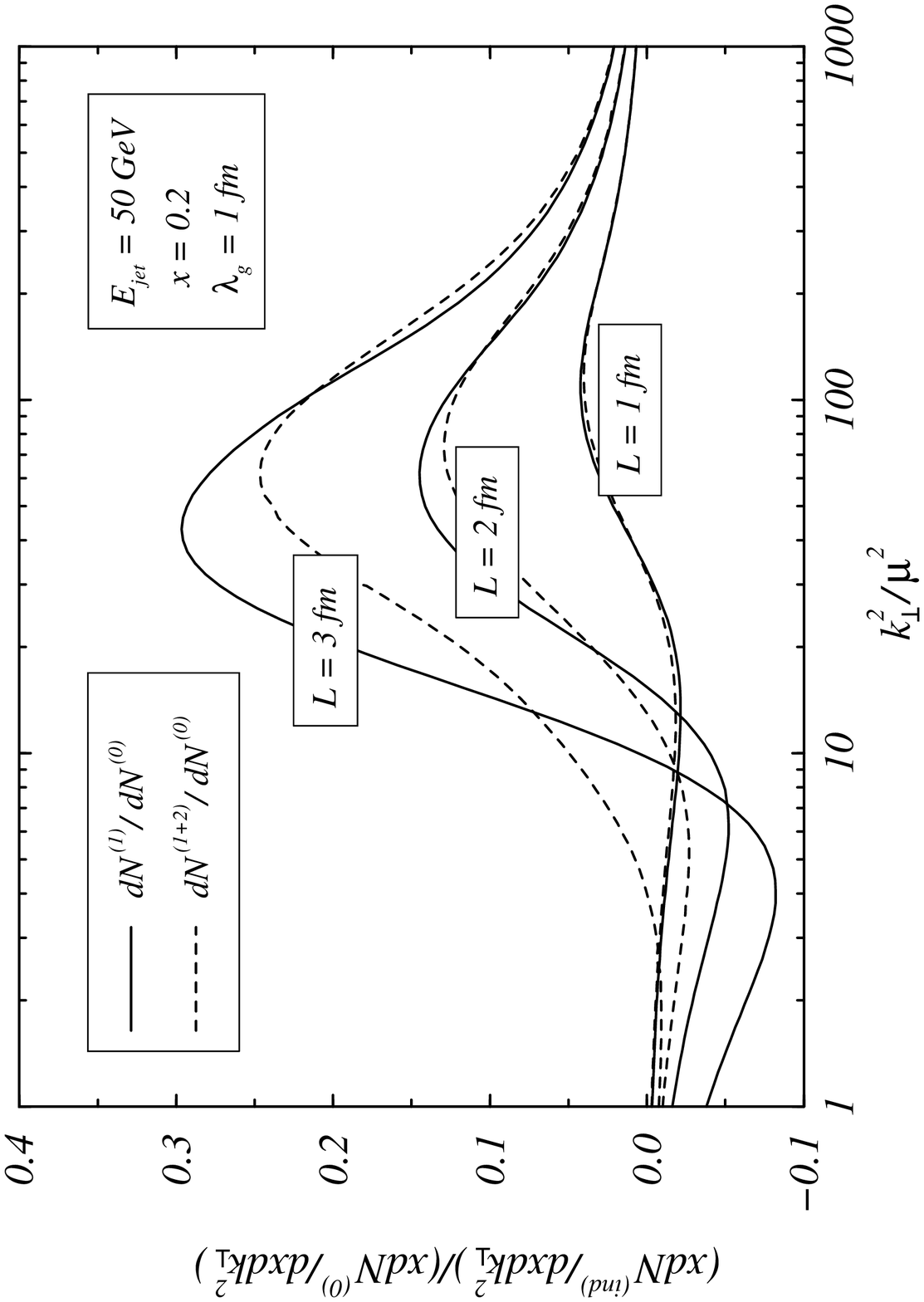}
\vspace{-2.6cm}
\end{center}
\begin{center}
\begin{minipage}[t]{8.6cm}
         { FIG. 1.} 
         {\small The medium induced gloun differential multiplicity 
          normalized by the factorization distribution eq.(\ref{hdist})
         is plotted vs. 
         ${\bf k}^2_\perp / \mu^2$ for opacity $L/\lambda_g=1,2,3$.
        Solid curves show first order (\ref{dnx1})
        and  dashed including second
        order in opacity (\ref{secord}) for a gluon with  $x=0.2$.
        (Quark jet of $E_{jet}=50$~GeV and $\mu=0.5$~GeV). }
\end{minipage}
\end{center}
\vspace{0.2cm}
\begin{center}
\vspace*{8.3cm}
\includegraphics{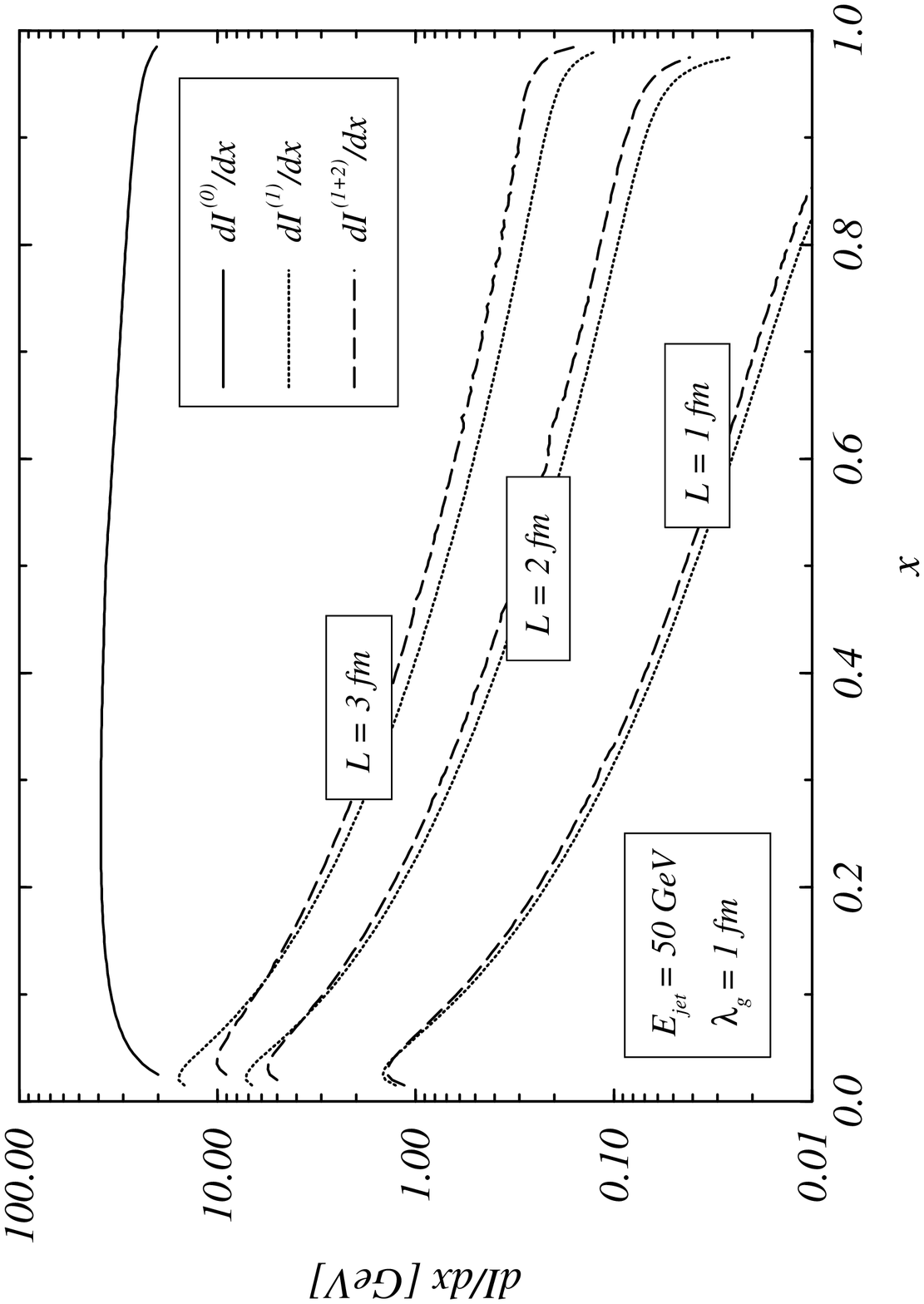}
\vspace{-2.0cm}
\end{center}
\begin{center}
\begin{minipage}[t]{8.4cm}  
      { FIG. 2.} {\small The contributions to the 
      induced radiation intensity to first and second order in opacity are
      compared to the medium independent intensity (\ref{di0}).
     The same conditions as in Fig.~1 are assumed.} 
\end{minipage}
\end{center}
\vspace{0.2cm}
\begin{equation}
\frac{dI^{(1)}}{dx} \approx \frac{C_R\alpha_s}{4} \,
\frac{ 1-x + \frac{x^2}{2} }{x} \, 
\frac{L^2\mu^2}{\lambda_g}\;.
\label{didx1app}
\end{equation} 
This formula  breaks down at both $x\rightarrow 0$ and
$x\rightarrow 1 $ because $|{\bf k_\perp}|_{\rm max}$ cannot be 
approximated by $\infty$ and because the small $x$ approximations used above
break down as $x\rightarrow 1$.

The total radiative energy loss is then given by
\begin{equation}
\Delta E^{(1)}=\frac{C_R\alpha_s}{N(E)}\, 
\frac{L^2\mu^2}{\lambda_g} \,\log \frac{E}{\mu}  \;\;,
\label{de1}
\end{equation}
\begin{center}
\vspace*{8.2cm}
\includegraphics{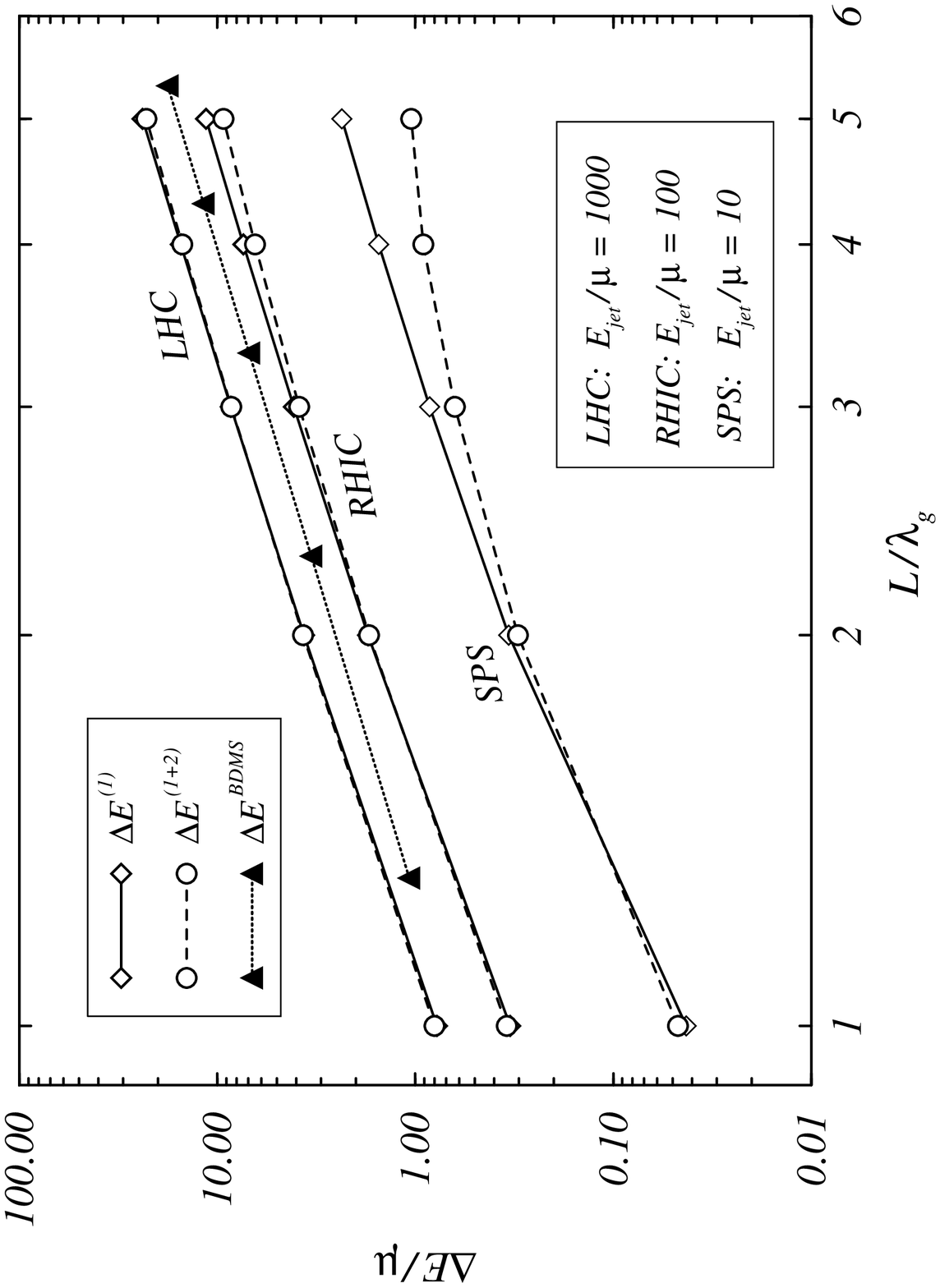}
\vspace{-2.2cm}
\end{center}
\begin{center}
\begin{minipage}[t]{8.6cm}  
        { FIG. 3.} {\small The radiated energy loss of a quark jet
        with energy $E_{jet}=5,50,500$~GeV  (at SPS, RHIC, LHC) 
        is plotted as a function of 
        the opacity $L/\lambda_g$.  ($\lambda_g=1$~fm, $\mu=0.5$~GeV).
        Solid curves show first order, while dashed curves show
        results up to second order in opacity. 
        The energy loss (solid triangles) 
         from Ref.~\cite{bdms} (with $\tilde{v}=2.5$)
        is shown for comparison.} 
\end{minipage}
\end{center}
\vspace{0.2cm}
with $N(E)=4$ if the kinematic bounds are ignored as in Eq.~(\ref{didx1app}).
In practice,  we emphasize that finite kinematic constraints
cause $N(E)$ to deviate considerably
from the asymptotic value 4. We find that 
$N(E)=7.3,\, 10.1,\, 24.4$  for $E=500,\, 50,\, 5$ GeV.
Together with the logarithmic
dependence of energy, these kinematic effects suppress greatly
the energy loss at lower (SPS) energies as seen in Fig.~3..
This is in contrast to the approximately energy independent result
in Ref.~\cite{bdms} where the finite kinematic bounds were neglected.

{\em  Conclusions.}
We calculated the effect of final state interactions on the induced gluon
differential distribution up to second order in opacity
for hard jets produced in nuclear reactions.
This work generalizes Ref.~\cite{glv1b} by taking into account
virtual corrections
to calculate inclusive rates and provides an complementary
 analytic approach to clarify
nonlinear jet quenching effects predicted
in Refs.~\cite{bdms_cone,bdms,zah}. 
The inclusive $xdN/dx d {\bf k}^2_\perp$, $dI/dx$,   
and $\Delta E$, was studied as a function of nuclear
thickness for jet energies in the SPS, RHIC and LHC
range.  One of our main results is  the demonstration
that the  second order  contribution (\ref{secord})
to  the integrated energy loss remains surprisingly 
small up to realistic nuclear 
opacities $L/\lambda_g\sim 5$ ( except at low SPS energies).
 The rapid convergence of the opacity expansion even for realistic
opacities results from the fact that the effective
expansion parameter is actually the product of the opacity and 
the gluon formation probability $L\mu^2/xE$ (see~(\ref{fapprox})).
The leading quadratic dependence
of the energy loss on nuclear thickness therefore
arises from the simple
first order term (\ref{dn1amps}) in this approach.
The detailed pattern of angular broadening and the $x$ dependence
is also dominated by the first order contribution.
We note that  our first order (effectively power-law) 
$|{\bf k_\perp}|$ and $x$ distributions,  however, differ
considerably from the Gaussian form obtained in the eikonal resummation
approach of~\cite{bdms_cone,bdms} that applies to thick targets.
Surprisingly, on the other hand, the magnitude of the 
integrated $\Delta E(L)$ is rather
similar at RHIC energies. 

At SPS energies kinematic effects
suppress greatly the energy loss relative to \cite{bdms}.
Our estimates provide a natural explanation for the absence of
jet quenching in $Pb+Pb$ at 160~AGeV that has been a puzzle up to
now~\cite{XNWA98,MGPL}. The short duration of the 
dense phase further limits the
effective opacity at the SPS.  A duration of $L/\lambda_g\sim 2$ for example
leads to a total energy loss only $\sim 100$ MeV, which is
much too small to be observable in soft multiple scattering 
background~\cite{MGPL}.  At RHIC energies, on the other hand, a significant
nonlinear (in $A$) pattern of suppression\cite{gptw,mgxw92} of high $p_\perp$
hadrons relative to scaled $pp$ data should be observable
to enable a direct test of non-abelian energy loss mechanisms in dense matter.
We note finally that the simplicity of the 
first and second order results (\ref{dn1amps},\ref{secord})
will make it possible to 
improve significantly  Monte Carlo event simulations
of  jet quenching~\cite{mgxw92}. 

\vspace{.3cm}

We thank A.~Mueller and U.~Wiedemann
for discussion on virtual corrections.
This work was supported by the 
DOE Research Grant under Contract No.
De-FG-02-93ER-40764, partly by the US-Hungarian Joint Fund No.652
and OTKA No. T029158.

[Note added in proof: U. Wiedemann informed us of an 
opacity expansion derived independently in~\cite{ursopac}.]

\vfill\eject
\end{multicols}
\end{document}